\documentclass[twocolumn,showpacs,preprintnumbers,amsmath,amssymb,aps]{revtex4}

\usepackage{graphicx}
\usepackage{dcolumn}
\usepackage{bm}
\usepackage[dvipdf]{hyperref}

\newcommand{\sro}{{$^{88}$}Sr}
\newcommand{\srs}{{$^{86}$}Sr}
\newcommand{\grad}{$^{\circ}\mathrm{C}$}

\newcommand{\tripl}{$^1$S$_0$-$^3$P$_1$}
\newcommand{\singl}{$^1$S$_0$-$^1$P$_1$}

\draft
\begin{document}
\title{Cooling and trapping of ultra-cold strontium isotopic mixtures}

\author{N. Poli}
\author{R. E. Drullinger}
\author{G. Ferrari}
\author{J. L\'eonard\footnote{Van der Waals-Zeeman Institute,
Amsterdam, The Netherlands\\ permanent address: IPCMS, Strasbourg, France.}}
\author{F. Sorrentino}
\author{G. M. Tino}
\email{Guglielmo.Tino@fi.infn.it}

\affiliation{Dipartimento di Fisica and LENS, Universit\`a di Firenze\\
Istituto Nazionale di Fisica Nucleare, Sezione di Firenze\\
Istituto Nazionale di Fisica della Materia, Unit\`a di Firenze\\
Polo Scientifico, 50019 Sesto Fiorentino, Italy}

\date{\today}

\begin{abstract}
We present the simultaneous cooling and trapping of an isotopic mixture in a
magneto-optical trap and we describe the transfer of the mixture into a conservative,
far-off resonant dipole trap. The mixture is prepared with a new technique that applies
to intermediate and heavy alkaline earth like atoms. In this work, \sro\ and \srs\ are
simultaneously loaded first into the magneto-optical trap operated on the \tripl\
spin-forbidden line at 689 nm, and then transferred into the dipole trap. We observe fast
inter-species thermalization in the dipole trap which allows one to set a lower bound on
the \sro-\srs\ elastic cross section.

\end{abstract}

\pacs{32.80.Pj, 42.62.Fi, 34.50.-s, 06.30.Ft. }
\maketitle

Recently laser cooled atomic strontium has been the subject of active researches in
various fields spanning from all-optical cooling towards quantum degeneracy for bosonic
and fermionic isotopes \cite{katori99,mukaiyama03}, cooling physics \cite{xu03,loftus04},
continuous atom laser \cite{katori00}, detection of ultra-narrow transitions
\cite{courtillot03,takamoto03,ferrari03,ido04}, multiple scattering \cite{bidel02}, and
collisional theory \cite{dereviankot03}.

\begin{figure}\begin{center}
\includegraphics[width=0.4\textwidth]{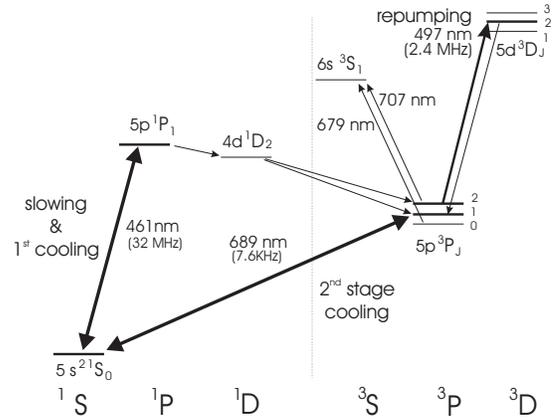}
\vspace{-2mm} \caption{\label{levl} Relevant energy levels and optical transitions for
laser cooling and trapping of strontium.}\end{center}
\end{figure}

As a result of the efficient second stage Doppler cooling, strontium is a promising
candidate to reach quantum degeneracy.  Second stage Doppler cooling \cite{katori99}
working on the \tripl\ intercombination transition (see figure~\ref{levl}) combined with
optical dipole trapping was proven effective to directly reach phase space densities of
$\rho=0.1$~\cite{Ido00}. Further increase of $\rho$ seems to be limited by light assisted
collisions and direct evaporative cooling was not reported so far. Alternative cooling
techniques are offered by mixtures of different atomic species~\cite{santos95} or
different isotopes~\cite{suptitz94}\cite{mewes99}. Mixtures offer a way to exploit
collisional physics not applicable in single species samples. They also offer additional
degrees of freedom such as sympathetic cooling in order to achieve the degenerate quantum
regime with atoms for which evaporative cooling is not efficient \cite{drullinger80}.

In this article we present the simultaneous cooling and trapping of isotopes of atomic
strontium in a magneto-optical trap (MOT)~\cite{raab87} and we describe the transfer of
the mixture into a conservative, far-off resonant dipole trap (FORT). The isotopic
mixture is prepared with a new technique based on sequentially loading different isotopes
into a magnetic trapped metastable state through the leakage of a MOT operating on a
single isotope. Afterwards the magnetically trapped mixture is optically pumped down to
the ground state for further cooling close to the recoil limit on the intercombination
\tripl\ transition. Then the mixture is loaded into a FORT. The advantage of this method
comparing with previous works based on simultaneous optical cooling and trapping of two
species \cite{suptitz94,mewes99}, is the drastic simplification of the laser source and
the elimination of possible interferences between the simultaneous loading into the MOTs.
For the sake of clarity we start by describing the production of a single-isotope sample,
then we explain how we extend the procedure to the production of isotopic mixtures.

In the experimental setup we use a thermal source of strontium atoms as described in
\cite{ferrari03}. The strontium is evaporated in an oven heated to 500\grad. The atomic
beam is collimated through a bundle of capillaries which limits the geometric divergence
to 50~mrad full width.  After the capillaries, the atomic beam brightness is increased by
a factor 4 with a transverse cooling stage realized with multipass beams~\cite{shimizu90}
in two orthogonal planes, with a frequency offset by 20~MHz to the red of the \singl\
transition ($\Gamma/2\pi$=32~MHz). After the collimation stage the atomic beam is slowed
to a few tens of m/s in a 30-cm long Zeeman-slower \cite{phillips82} based on a two stage
tapered coil with zero crossing magnetic field and a counter-propagating laser beam tuned
15~$\Gamma$ to the red of the \singl\ transition. The light beam is typically 18~mW, has
a 7 mm $e^{-2}$ radius at the MOT and is focused  on the capillaries, 85~cm away from the
MOT. The slowed atoms are then further cooled and trapped in a MOT working on the \singl\
transition red detuned by 40~MHz from the resonance. The MOT is formed by three
retro-reflected beams with a $e^{-2}$ radius of 5~mm crossing almost orthogonally, with
the vertical beam collinear with the magnetic quadrupole axis of an anti-Helmholz pair of
coils. The total 461~nm light incident on the MOT region amounts to 46~mW/cm$^2$, and the
magnetic gradient is 56 G/cm. The atom number, spatial density and temperature are
derived in time-of-flight by imaging on a CCD camera the atomic absorption of a 70 $\mu$s
probe beam resonant on the \singl transition.  The 461~nm light source for the first
cooling stage, as described in \cite{ferrari04}, is based on frequency doubling of a
semiconductor laser. It routinely delivers 150~mW at 461~nm in a 4~MHz linewidth FWHM.
The source is frequency stabilized with saturated spectroscopy on a strontium vapor
cell~\cite{ExplicationHeatpipe}.

After the initial capture in the blue MOT, and following the approach described in
\cite{katori99}, the atoms are further cooled in a MOT operating on the \tripl\
intercombination line at 689 nm. The 689~nm light is provided by a master-slave
semiconductor laser system which is frequency narrowed by locking to an optical cavity
which is, in turn, steered to the \tripl\ transition by saturation spectroscopy in a
second strontium heatpipe~\cite{ferrari03}. The red beam is mode-matched to the 461~nm
MOT beam and it is overlapped to it on a dichroic mirror. From this point the two beams
share the same broad-band optics up to the atoms.

\begin{figure}\begin{center}
\vspace{-6mm}
\includegraphics[width=0.4\textwidth]{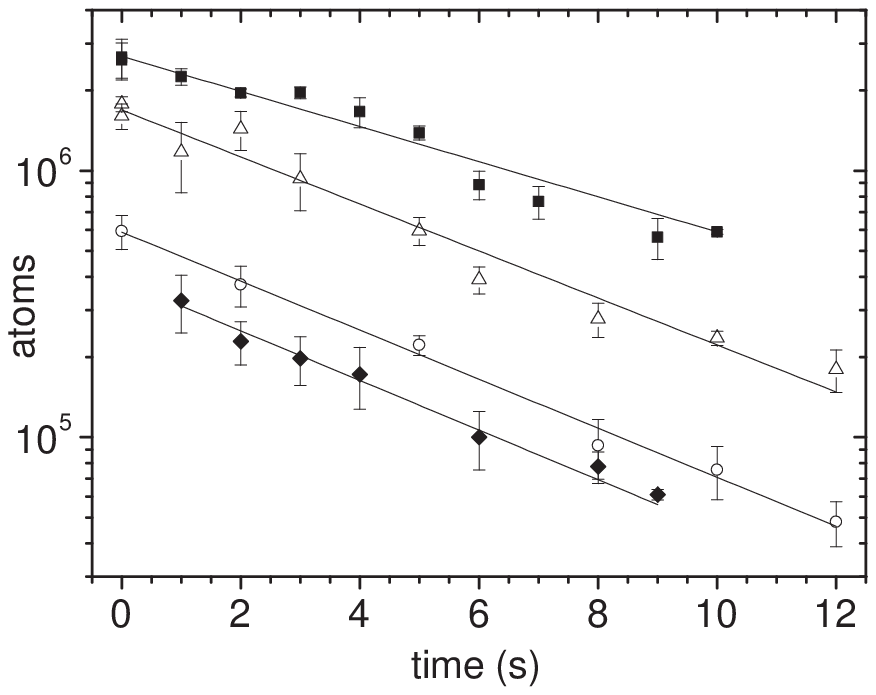}
\vspace{-7mm} \caption{\label{MetastableDecay} Metastable $^3$P$_2$ state decay when
trapped in a 56 G/cm magnetic quadrupole. The data are taken for the individual isotopes
with the \singl\ MOT switched off (circles: \srs, lifetime $\tau$= 4.7~s. Squares: \sro,
$\tau$= 6.6~s), and with the \singl\ MOT operating on the undetected isotope (diamonds:
\srs\ with \sro\ MOT, $\tau$= 4.7~s. Triangles: \sro\ with \srs\ MOT, $\tau$= 4.9~s). The
measurements are taken after red MOT recapture and the $e^{-1}$ lifetimes are reported.}
\end{center}
\end{figure}

As discussed in the following, the atoms leak from the blue MOT to the metastable
$^3$P$_2$ due to the $2\times 10^{-5}$ branching ratio of the $^1$P$_1$ excited state to
the low-lying $^1$D$_2$ and the subsequent decay to the $^3$P$_2$ state. This loss
channel is eliminated with a single optical pumping process which involves the
$^3$P$_2-^3$D$_2$ transition at 497~nm. For this purpose we developed an anti-reflection
coated laser-diode stabilized in the Littrow ECL configuration which delivers 25 mW at
994 nm. This light is frequency doubled on a 17 mm long KNbO$_3$ crystal placed in a
resonant cavity for improved conversion efficiency resulting in 4 mW at 497 nm. The
KNbO$_3$ is b-cut for non-critical phase matching at a temperature of 54\grad. The 497~nm
beam is sent to the MOT region with a 1.5 cm $e^{-2}$ diameter, it is retroreflected, and
its frequency is tuned in order to maximize the blue MOT fluorescence.

Without the 497~nm repumper the blue MOT lifetime is limited to 10~ms and typically we
trap 10$^7$ \sro\ atoms at 3~mK. By applying the green repumper the blue MOT lifetime
increases by more than one order of magnitude, resulting in up to $3\times10^8$ atoms
loaded into the \sro\ blue MOT at a density of $6\times 10^9$~cm$^{-3}$. These values are
comparable to the alternative optical pumping scheme~\cite{vogel99} in which two lasers
at 707 and 679 nm couple the $5p\,^3$P$_2$ and the $5p\,^3$P$_0$ states to the
$6s\,^3$S$_1$ state (see figure \ref{levl}).

As described in \cite{katori99}, the transfer from the blue to the red MOT is done in two
steps: For the first 200 ms the red laser is frequency modulated at 50~kHz, spanning
2~MHz (FM red MOT), to recapture the atoms moving faster than the capture velocity on the
\tripl transition. In the following 40 ms the red laser is set single frequency at the
intensity that determines the desired final temperature of the sample. Working at 350~kHz
below resonance and reducing the total light intensity on the MOT to 70~$\mu$W/cm$^2$, we
obtain about 10\% of the initial atoms from the blue MOT cooled below 1~$\mu$K.

For simultaneous trapping of multiple isotopic samples previous experiments employed
laser sources delivering the necessary frequency components for each isotope involved
\cite{suptitz94,mewes99}. This approach in the case of the blue MOT may be difficult to
apply because of the complexity of the laser sources, the limited optical access, the
frequency offsets involved, and the limited laser power. An alterative solution is
presented by the use of the magnetically trapped $^3$P$_2$ state as a dark atom
reservoir~\cite{stuhler01}. During the blue MOT phase without the repumper, the small
branching ratio of the excited $^1$P$_1$ state towards the metastable $^3$P$_2$ state,
provides a continuous loading of atoms into the potential given by the MOT's magnetic
quadrupole \cite{nagel03}. Figure \ref{MetastableDecay} reports the lifetime of the
magnetically trapped metastable isotopes. Then, using the same blue source, one can
sequentially load different isotopes into the magnetic potential just by stepping the
laser frequency to the different resonances.

\begin{figure}\begin{center}
\includegraphics[width=0.45\textwidth]{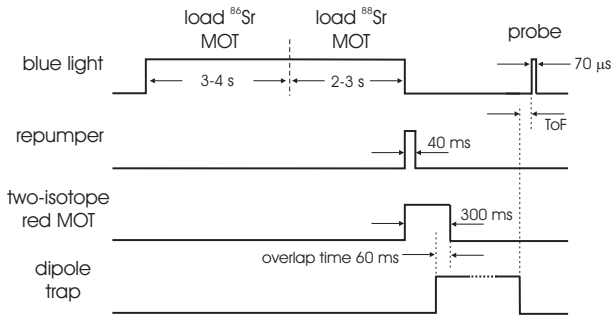}
\caption{\label{seq} Time sequence for simultaneous trapping of \sro\ and \srs\ in an
 optical dipole trap. MOT: magneto-optical trap. ToF: time of flight.}\end{center}
\end{figure}

The trapping sequence for collection of a \sro -\srs\ mixture is reported in
Figure~\ref{seq}. We start collecting \srs\ for 3~s, then we tune the blue laser on
resonance to \sro\ for 2~s. Once the isotopic mixture is prepared in the $^3$P$_2$ state,
the blue light is switched off, and the FM red MOT is switched on as well as the
repumping beam. The isotopic shift on the repumping transition is smaller than the
resonance width of the $^3$P$_2$-$^3$D$_2$ transition observed on the blue MOT
fluorescence, which results in efficient, simultaneous optical pumping of \sro ~and \srs
~on a timescale short with respect to red MOT capture time \cite{ExplicationDRepumping}.
The loading of a single isotope into the magnetic potential was already described in
\cite{nagel03} and we did not observe significant differences in the behavior when
loading two isotopes. Figure \ref{MetastableDecay} shows the measurement of the lifetime
for each isotope, both when individually trapped and in presence of the blue MOT working
on the other isotope. All the lifetimes are on the order of 5~s, close to the background
pressure limited lifetime of 8~s.

The laser source for the operation of the two-isotope red MOT is composed of two slave
lasers injected from the same frequency-stabilized master with a frequency offset
corresponding to the isotopic shift of 163 817.3 kHz \cite{ferrari03}. Subsequently, the
frequency and the intensity of the two beams are controlled by double pass AOMs driven by
the same RF, the beams are superimposed on a polarizing beam splitter, and then they are
overlapped to the blue MOT beams as described previously.

Comparing the two-isotope red MOT with the single isotope one, with the same atom numbers
we do not observe any effect in the transfer efficiency and final temperature due to the
presence of the second isotope. In this way, we obtain samples with up to $10^7$ ($10^6$)
atoms of \sro\ (\srs) at a temperature 2~$\mu$K (1~$\mu$K). We attribute the difference
in the loading with respect to the natural abundances to minor differences in the red MOT
parameters. By varying the order of loading and the loading times of the two isotopes we
can vary almost arbitrarily the final ratio of populations.

\begin{figure}\begin{center}
\vspace{-7mm}
\includegraphics[width=0.4\textwidth]{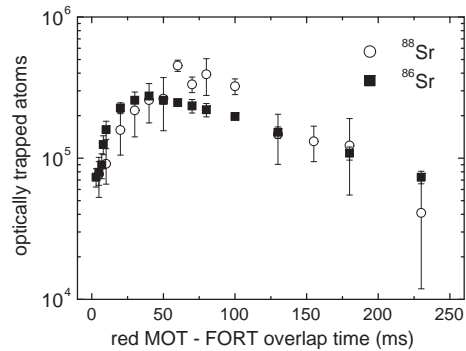}
\vspace{-7mm} \caption{\label{load} Transfer from the intercombination red MOT to the
optical dipole trap with different overlap time between the final red MOT and the FORT.
The \srs\ and \sro\ mixture was prepared in order to load about twice \sro\ than \srs\
atoms in the final red MOT. }
\end{center}
\end{figure}

The cloud is then transferred into a FORT realized by focusing a far detuned 922 nm laser
beam on the center of the trap. For this application we employ a semiconductor tapered
amplifier seeded from the same 922 nm source used for the 461 nm laser. At this
wavelength, setting the optical field polarization orthogonal to the magnetic field, the
differential stark shift between the $^1$S$_0$ and $^3$P$_1$ levels is small with respect
to the transition linewidth, allowing precision spectroscopy and laser cooling even on
optically confined atoms \cite{ido03}. The light from the amplifier is spatially filtered
through a polarization-maintaining single-mode optical fiber and is focused on the atoms
to a waist of 15~$\mu$m. At maximum power, we measure the radial oscillation frequency
$\nu_{\rm rad}$=2~kHz and the axial oscillation frequency $\nu_{\rm ax}$=26~Hz. These
values are consistent with our peak intensity of 160 KW/cm$^2$ which corresponds to a
trap depth of 90~$\mu$K.

As shown in figure \ref{seq}, the transfer between the \tripl\ MOT and the FORT is done
by adding the 922~nm light 60~ms before switching off the MOT. Figure \ref{load} shows
the transfer of the mixture from the red MOT to the FORT as a function of the overlap
time. The loading dynamics of \srs\ is slightly faster than for \sro\, while the losses
at overlap time longer than 100~ms are attributed to the balance of light assisted
collisions \cite{katori99}. The maximum transfer efficiency of 40~\% is observed by
overlapping the FORT beam waist on the red MOT, and then displacing the the FORT waist
500~$\mu$m in the direction of the FORT beam. This displacement corresponds to 2/3 of
Rayleigh range, or equivalently 2 red MOT $e^{-2}$ radii. The effects of this
displacement are the increase of the overlap volume between the final red MOT and the
FORT, and the reduction of the light assisted collisional losses at the bottom of the
dipole potential. Finally we load into the FORT up to 3 10$^5$ (10$^6$) atoms typically
at 15~$\mu$K (20~$\mu$K) for the individual $^{86}$Sr ($^{88}$Sr) isotope. The difference
in the final temperature is in agreement with the density dependent heating reported in
\cite{katori99}. When loading the isotopic mixture we do not observe any reduction of the
transfer efficiency from the red MOT into the FORT. Surprisingly in the optical trap we
always find thermal equilibrium among the two species. The inter-species thermalization
is fast on the timescale of the loading into the FORT and allows us to set a lower bound
for the inter-species elastic cross section $\sigma > 3.5\times 10^{-12}$~cm$^{2}$.
Equivalently, the modulus of the inter-species scattering length is
$|a|=\sqrt{\frac{\sigma}{8\pi}}>70$ Bohr radii.

In conclusion we presented the simultaneous cooling and trapping of the \sro\ and \srs\
isotopes into a MOT operating on the spin-forbidden \tripl\ transition, and we described
the transfer of the mixture into a conservative, far-off resonant, dipole trap working
near the ``{\it magic wavelength}'' for the \tripl\ transition \cite{ido03}. The method,
which is demonstrated with strontium isotopes, can be applied to all the atoms with
similar level structure and transitions, to load an arbitrarily large number of isotopes.
The realization of the strontium isotopic mixture and the observed strong inter-species
thermalization will allow the implementation of additional cooling mechanisms based
sympathetic cooling. This seems very promising towards the realization of a Bose-Einstein
condensate of an alkali-earth atom.

We acknowledge fruitful discussions with M. Prevedelli, J. L. Hall, and C. Salomon, and
contributions in the early stages of the experiment from P. Roulleau. We thank R.
Ballerini, M. De Pas, M. Giuntini, and A. Hajeb for technical assistance. J.L.
acknowledges support from the EU network on Cold Quantum Gases (HPRN-CT-2000-00125). This
work was supported by Agenzia Spaziale Italiana, Ente Cassa di Risparmio di Firenze,
MIUR-COFIN 2002, Istituto Nazionale Fisica della Materia, Istituto Nazionale di Fisica
Nucleare     and LENS.

\end{document}